\documentstyle[12pt]{article}

\def\be{\begin{equation}}
\def\ee{\end{equation}}

\begin{document}
\title{Towards common origin of supersymmetry breaking,  compositeness,
and gauge mediation}
\author{{\em S.~L.~Dubovsky\footnote{{\em e-mail:} sergd@ms2.inr.ac.ru}, 
             D.~S.~Gorbunov\footnote{{\em e-mail:} gorby@ms2.inr.ac.ru},
         and S.~V.~Troitsky\footnote{{\em e-mail:}    st@ms2.inr.ac.ru}}\\
   {\small {\em Institute for Nuclear Research of the 
  Russian Academy of Sciences,}}\\
  {\small {\em 60th October Anniversary 
  prospect, 7a, Moscow 117312, Russia.}}
  }
\date{December 16, 1997}
\maketitle
\begin{abstract}
We present a toy model which has the following properties: i) it is
strongly coupled at a certain scale $\Lambda_S$ and breaks
supersymmetry at this scale; ii) matter superfields of (one generation
of) the Standard Model are effective low energy degrees of freedom of
this strongly coupled theory; iii) direct gauge mediation of
supersymmetry breaking occurs due to messenger superfields that are
also effective low energy degrees of freedom.  We discuss the way to
generalize this approach to realistic models with three generations.
These models might be favorable from the point of view of string
theory compactifications.
\end{abstract}

{\bf 1.} The purpose of this letter is to 
propose an approach that incorporates in an economical way both
supersymmetry breaking and its mediation to the
visible sector.  We present a toy model in which 
the Standard Model matter 
fields of one generation are
composite and appear as low energy effective degrees of freedom of
another theory which breaks supersymmetry at strong coupling. 
A subgroup of the flavor symmetry group 
of the strongly coupled theory 
is gauged and
identifyed with the gauge group of the Standard Model.
We begin with our motivations, then
describe the toy model, and finally discuss possible ways to generalize
the approach in order to construct realistic models with 
three generations.

The interest in realistic models in which supersymmetry is broken
dynamically in a secluded sector \cite{ADS-ph} and mediated to
observable fields by means of gauge interactions \cite{gm-old} has
been renewed in recent years \cite{Dine1-2-3} (see
Refs.~\cite{Skiba,Kolda} for reviews and references) as novel
techniques of analysing strongly coupled $N=1$ supersymmetric gauge
theories have been developed \cite{Seiberg1,Seiberg2}. The advantages
of gauge-mediated supersymmetry breaking are its predictive power and
natural explanation of the suppression of flavor-changing processes.
Until now, one of the main drawbacks of this approach is complicated
structure of realistic models. As a rule, more simple structures 
appear in the models with so-called ``direct gauge mediation'' 
(see Ref.\cite{best-model} for one of elegant examples). 
It is desirable that apparently
different phenomena -- supersymmetry breaking, appearance 
of both
matter content of the Standard Model and messenger superfields -- are
manifestations of one and the same mechanism. Here we propose how
such a mechanism may emerge due to strong coupling dynamics of
supersymmetric gauge theories. Several attempts
to construct realistic models exploiting strong coupling dynamics 
have been made
but compositeness, supersymmetry breaking and its mediation to visible 
sector appeared from completely
different mechanisms
(see, e.g., Ref.\cite{compos}). Though our toy model is far from being
viable, we hope that its main features are common to realistic models
exploiting the same mechanism.

The idea of the approach discussed here --- to consider the Standard
Model group as a gauged flavor symmetry of a strongly coupled group
that dynamically breaks supersymmetry (DSB group) --- is relatively
old and has faced phenomenological problems (see, e.g.,
Ref.~\cite{ADS-ph}). The key novelty suggested in this paper is to identify
both the Standard Model matter and messengers with effective 
low energy degrees of freedom of DSB
theory instead of introducing them as fundamental fields. 
In this approach, the 
direct mediation scheme appears in a natural way.

{\bf 2.}  Let us turn now to the toy model exhibiting these
features.  The main ingredient is $SU(5)_S$ gauge theory with matter
in one chiral (antisymmetric tensor $A$ plus antifundamental
$\bar{Q}_6$) and five vector-like (fundamental $Q_i$ plus antifundamental
$\bar Q_i$, $i=1,\dots,5$) generations.  Due to the 
presence of five fundamentals and six
antifundamentals,  the global symmetry group besides $U(1)$ factors
contains $SU(5)\times  SU(6)$ flavor symmetry. We embed the Standard
Model gauge group into the $SU(5)_W$ subgroup of this flavor
symmetry group as follows: only fields from vector-like generations are
charged under the  Standard Model group, and $SU(5)_W$ is the vector
subgroup of the $SU(5)\times SU(5)$ flavor symmetry of $Q_i$ and $\bar Q_i$. 
So, the matter content of the fundamental high-energy theory is:

\begin{center}
\begin{tabular}{ccc}
           &$SU(5)_S$ &$SU(5)_W$ \\
$A$        &  10      & 1        \\
$\bar Q_6$ & $\bar 5$ & 1        \\
$\bar Q$ & $\bar 5$ & 5        \\
$ Q$     & 5        & $\bar 5$ \\
\end{tabular}
\end{center}

Both $SU(5)_S$ and the Standard Model are asymptotically free. We 
consider the case when $SU(5)_S$ becomes strongly coupled at a
relatively high scale $\Lambda_S$. At this scale the Standard Model group
is
still weakly coupled and unbroken.
The dynamics at energy scales below $\Lambda_S$ 
is described by an effective theory which is dual to $SU(5)_S$ 
in the sense of Ref.~\cite{Seiberg2}. 
The dual theory to $SU(5)$ with one antisymmetric tensor has
been constructed in Ref.~\cite{Pouliot}. In our case the dual gauge
group is $SU(2)\times SU(2)$, and the matter content of the 
low energy effective
theory is:
$$
\begin{array}{cccc}
         &  SU(2)_1 &SU(2)_2   & SU(5)_W   \\
x        &     2    & 2        &  1        \\
p        &     2    & 1        &  1        \\
\bar a   &     1    & 1        &  1        \\
l_6      &     1    & 2        &  1        \\
l        &     1    & 2        &  \bar 5   \\
\bar q   &     2    & 1        &  5        \\
M_6      &     1    & 1        &  \bar 5   \\
M        &     1    & 1        &  1+24     \\
H_6      &     1    & 1        &  5        \\
H        &     1    & 1        &  10       \\
B_1      &     1    & 1        &  \bar 5   \\
\end{array}
$$
We use the notations of Ref.~\cite{Pouliot} but write explicitly the
degrees of freedom carrying the sixth index of $SU(6)$ flavor subgroup
of the original theory. 
The dual theory has a superpotential (for simplicity
all indices including the sixth are contracted),
\begin{equation}
W={1\over\Lambda_S^2}M\bar q lx+{1\over\Lambda_S^2}Hll+
{1\over\Lambda_S^2}B_1 p\bar q+\bar a x^2.
\label{spot}
\end{equation}

When explicit mass term for fundamental flavors, $mQ_i\bar Q_i$, is
added, $m\ll\Lambda_S$, the low energy potential (\ref{spot})
acquires an extra term $mM^{(1)}$, where $M^{(1)}$ is the $SU(5)_{W}$
singlet, and  supersymmetry  breaks down. To see this,
consider the effect of this perturbation on the low energy effective
superpotential. The term $mM^{(1)}$ for $SU(5)_W$ forces the $F$-term
equations to be inconsistent with each other. Indeed, 
$$
F_M^{(1)}={1\over\Lambda_S}\bar q l x+ m\Lambda_S=0,
$$
$$
F_{\bar a}=x^2=0
$$
have common solutions only for $m=0$.~\footnote{Note that in the
general case of arbitrary number of flavors in the fundamental theory
supersymmetry breaking requires adding not only the mass term but also
Yukawa-like coupling~\cite{Pouliot}.  The reason is that the field
$\bar a$ transforms as an antisymmetric tensor under the dual gauge
group in general case while for $SU(2)$ group it is just a singlet.}
The (non-supersymmetric) flat directions of the model are raised by
means of perturbative corrections to K\"ahler potential. It is
straightforward to see that there is a supersymmetry breaking vacuum
with $F_M=m\Lambda_S$ and all expectation values of fields equal to
zero,
so the full low energy gauge group is unbroken (there may be other vacua
where fields charged under $SU(5)_W$ gauge group get vacuum
expectation values via the inverse hierarchy mechanism, but
they might occur only at field values larger than $\Lambda_S$ where
the dual description is not valid). 

Let us discuss the effective theory in more detail.  Fields charged
only under $SU(5)_W$ are $H$, $B_1$, $H_6$, $M_6$.  They fall into one
chiral generation, ($10+\bar 5$), and one vector-like flavor, ($5+\bar
5$) --- just the matter representations involved in the simplest GUT
extension of the Standard Model.  The fields $l$ and $\bar q$ are
charged under both $SU(5)_W$ and $SU(2)_1\times SU(2)_2$ groups, and
so may serve as two sets of ($5+\bar 5$) messengers of supersymmetry
breaking.  Scalar components of $\bar q$ and $l$ get masses in one
loop due to nonzero $F_M$ (the relevant interactions involve the first
term in the superpotential).  Superpartners of quarks and leptons of
the Standard Model get masses through interactions with messengers.
This is not sufficient, however, to obtain non-zero masses of gauginos.
One needs that some fermions are massive, or some scalar field
developes a non-zero vacuum expectation value.  The only way to
provide these features is to add the only explicit mass term
consistent with all symmetries of the model, $kM^2$. This term leads
to appearence of a supersymmetric vacuum state at very large value of
singlet $M$, and to the shift of the supersymmetry breaking vacuum under
consideration away from the origin, bringing the vacuum expectation
value  $M\sim-k\Lambda_S^4/m$ to the singlet. Now gauginos get non-zero
masses through the interactions with messengers. Messenger fermions,
as well as higgsinos, are,
however, still massless. We need an additional
mechanism to provide these fermions with mass, which cannot be read
out from the effective theory discussed above, and this is one of the
most serious problems with our scenario.  Quarks and leptons
(fermions from $10+\bar 5$) remain massless due to their chirality.
 
To generate the term $kM^2$ in the low-energy theory, we suppose (in
the spirit of Ref.~\cite{Randall}) that high energy scale effects
generate non-renormalizable term ${1\over\mu}(Q\bar Q)^2$ in the
fundamental theory, where $\mu$ is some large mass scale where new
physics becomes essential.  Note that the mass term for $Q$ breaks
explicitly the flavor $SU(6)\times SU(5)$ group, as well as one linear
combination of the three global $U(1)$ symmetries of the theory listed
in Ref.~\cite{Pouliot}. This symmetry breaking may be spontaneous, if
 $m$
and $\mu$ originate from the expectation values of some
fields transforming non-trivially
under the flavor group.  

So, our model includes a secluded DSB sector,
simple direct gauge mediation mechanism, and the Standard Model
GUT-like weakly coupled theory at low energies.  
Though we discussed the physics in terms of $SU(5)_W$, the picture is
similar when the gauge group is $SU(3)\times SU(2)\times U(1)$.

One of the attractive features of the mechanism is that perturbative
unification of couplings at least in one loop is not spoiled. Indeed,
upon breaking $SU(5)_W$ down to the Standard Model product gauge group
in the usual way (by making use of a heavy adjoint Higgs field), 
running of the coupling constants will exhibit a 
threshold at $\Lambda_S$ where the low energy degrees of
freedom are replaced by the fundamental states. Since all of them 
 fall
in the complete GUT multiplets, 
the fact of unification remains intact and only unification scale and
corresponding value of couplings slightly change
(adding complete GUT multiplets leads
to a simultaneous shift in the three beta 
function coefficients). 

{\bf 3.}  Let us now discuss the ways to generalize the mechanism
presented above to construct more realistic models. To have three chiral
generations in the low energy $SU(5)_W$, one has, of course, to
enlarge the matter content of the fundamental DSB theory. If the
numbers of chiral generations in the high-energy DSB theory and in the
low-energy $SU(5)_W$ were necessarily  equal (as in the example
presented above), 
the scheme would be just a toy because $SU(5)_S$ breaks supersymmetry
only with one or two (10+$\bar 5$)
generations~\cite{ADS-ph,ADS-two-gen}. With three chiral generations,
even adding the superpotential cannot lift all flat
directions~\cite{ADS-ph}. Fortunately,  the
equality between numbers of chiral generations generally does not hold
even in the $SU(5)_S\times
SU(5)_W$ case. To show this fact by construction, let us exploit once
again the duality of Ref.~\cite{Pouliot} and consider $SU(5)_S$ with
one antisymmetric tensor, ten fundamentals and eleven
antifundamentals. Now we divide ten (5+$\bar 5$)-plets of $SU(5)_S$
into two 
sets of fundamentals and antifundamentals
 of gauged $SU(5)_W$ subgroup of
$SU(10)\times SU(11)$ global symmetry.  Though the dual model and its
matter content are too complicated to be discussed or even described
here in detail, we point out that $SU(5)_W$ group has three antisymmetric
tensor multiplets in the low energy description.

The possibility to have three chiral generations in the Standard Model
while only one is present in the high-energy theory is very
intriguing. In this context let us recall the problem of reconciling
the string compactification predictions and conventional Grand
Unification (see Ref.~\cite{Dienes} for a review): it is hardly
possible to obtain three chiral generations {\it and} Grand
Unification Higgs in the low energy limit of the string theory.  The
usual way to overcome this difficulty is to modify the unification
mechanism; in our case one may give up the requirement of three chiral
generations thus remaining open the possibility to use the conventional
scenario of unification.  Let us also point out that in this scenario,
the high energy theory is expected to contain a lot of matter, and the
perturbative unification of couplings is likely to be replaced by
phenomenologically viable non-perturbative one~\cite{strong-unif}
which is highly preferred from the string theory pont of
view~\cite{strong-string}.

To summarize, we presented a  simple mechanism of direct gauge
mediation of supersymmetry breaking. Both the Standard Model matter
and messengers are
effective low-energy degrees of freedom of the DSB
group. The mechanism does not spoil the unification of couplings
and opens up a possibility to construct realistic models
consistent with string compactification predictions. 

The authors are indebted to M.V.Libanov 
and V.A.Rubakov
for numerous helpful 
discussions. The work is supported in part by Russian Foundation for 
Basic Research grant 96-02-17449a. The work of 
S.T. is supported in part by the U.S.\
Civilian Research and Development Foundation for Independent States of
FSU (CRDF) Award No.~RP1-187 and by INTAS grant 94-2352. 

{\bf Note added:}
When this paper was completed, we received the preprint, Ref.~\cite{new},
where common mechanism of dynamical supersymmetry breaking and
compositeness is also discussed. The model of Ref.~\cite{new} is,
however, different from the one considered here.

\end{document}